\documentclass[iop,apjl]{emulateapj}
\usepackage{amsmath}
\usepackage{url}
\usepackage{natbib}
\usepackage[citecolor=blue]{hyperref}
\hypersetup{
    colorlinks=true,
    linkcolor=blue,
    filecolor=magenta,      
    urlcolor=blue,
}

\newcommand{\solar}{$_{\odot}$}
\newcommand{\wsqmsr}{W\,m$^{-2}$\,sr$^{-1}$}

\newcommand{\hcop}{HCO$^+$}
\newcommand{\httco}{H$^{13}$CO$^+$}

\newcommand{\joz}{$J$=1$\rightarrow$0}

\newcommand{\degree}{$^{\circ}$}

\newcommand{\nhtwo}{$N_{\rm H_2}$}
\newcommand{\ost}{[O\,\textsc{i}]$\lambda63~\mu$m}
\newcommand{\oee}{[O\,\textsc{iii}]$\lambda88~\mu$m}
\newcommand{\ooff}{[O\,\textsc{i}]$\lambda145~\mu$m}
\newcommand{\cofe}{[C\,\textsc{ii}]$\lambda158~\mu$m}

\shorttitle{Massive Young Protostars in BYF\,73}
\shortauthors{Pitts et al.}

\begin{document}

\title{Gemini, SOFIA, and ATCA reveal very young, massive protostars \\ in the collapsing molecular cloud BYF\,73}

%% Use \author, \affil, and the \and command to format author and affiliation information. 
\author{Rebecca L. Pitts\altaffilmark{1}, Peter J. Barnes\altaffilmark{1,2}, Stuart D. Ryder\altaffilmark{3,4}, and Dan Li\altaffilmark{5} \\
% William J. Schap III\altaffilmark{1}, and Eric Pantin\altaffilmark{5}
}
\email{rlpitts@ufl.edu}
%\and
%\author{}
%\affil{}

\altaffiltext{1}{Astronomy Department, University of Florida, P.O. Box 112055, Gainesville, FL 32611, USA}
\altaffiltext{2}{School of Science and Technology, University of New England, Armidale NSW 2351, Australia}
\altaffiltext{3} {Department of Physics and Astronomy, Macquarie University, NSW 2109, Australia}
\altaffiltext{4}{Australian Astronomical Observatory, 105 Delhi Road, North Ryde, NSW 2113, Australia}
\altaffiltext{5}{National Optical Astronomy Observatory, 950 North Cherry Avenue, Tucson, AZ 85719, USA}
%\altaffiltext{5}{CEA Saclay,}
%\altaffiltext{3}{Astronomy Department, University of Wisconsin, 475 North Charter St., Madison, WI 53706, USA}
%\altaffiltext{4}{College of Optical Sciences, University of Arizona, 1630 E. University Blvd., P.O. Box 210094, Tucson, AZ 85721, USA}
%\altaffiltext{4}{National Astronomical Observatory of Japan, Chile Observatory, 2-21-1 Osawa, Mitaka, Tokyo 181-8588, Japan}
%\altaffiltext{6}{Society of Fellows, Harvard University}
%\altaffiltext{7}{Patron, Alonso's Bar and Grill}
%\altaffiltext{8}{All-round Mensch}

\begin{abstract}
%{\color{red}Check all numbers!}
We present multi-wavelength data on the globally infalling molecular cloud/protostellar cluster BYF\,73.  These include new far-infrared (FIR) spectral line and continuum data from the Stratospheric Observatory for Infrared Astronomy's (SOFIA's) Far-Infrared Field-Imaging Line Spectrometer (FIFI-LS), mid-infrared (MIR) observations with the Thermal-Region Camera Spectrograph (T-ReCS) camera on Gemini-South, and 3mm continuum data from the Australia Telescope Compact Array (ATCA), plus archival data from {\em Spitzer}/Infrared Array Camera (IRAC), and {\em Herschel}/Photodetecting Array Camera and Spectrometer (PACS) and Spectral and Photometric Imaging Receiver (SPIRE).  The FIFI-LS spectroscopy in \ost, \oee, \ooff, and \cofe\ highlights different gas environments in and between the dense molecular cloud and H\textsc{ii} region.  The photo-dissociation region (PDR) between the cloud and H\textsc{ii} region is best traced by \ooff\ and may have density $>$10$^{10}$\,m$^{-3}$, but the observed $\lambda145\mu$m/$\lambda63\mu$m and $\lambda63\mu$m/$\lambda158\mu$m line ratios in the densest gas are well outside model values.  The H\textsc{ii} region is well-traced by [C\,{\sc ii}], with the $\lambda158\mu$m/$\lambda145\mu$m line ratio, indicating a density of 10$^{8.5}$\,m$^{-3}$ and a relatively weak ionizing radiation field, 1.5 $\lesssim$ log$(G/G_0)\lesssim$ 2.  The T-ReCS data reveal eight protostellar objects in the cloud, of which six appear deeply embedded ($A_V$ $>$ 30$^m$ or more) near the cloud's center.  MIR\,2 has the most massive core at $\sim$240\,$M$\solar, more massive than all the others combined by up to tenfold, with no obvious gas outflow, negligible cooling line emission, and $\sim$3--8\% of its 4.7$\times$10$^3$\,$L$\solar\ luminosity originating from the release of gravitational potential energy.  MIR\,2's dynamical age may be as little as 7000\,years.  This fact, and the cloud's total embedded stellar mass being far less than its gas mass, confirm BYF\,73's relatively early stage of evolution.
\end{abstract}

%% The macro also takes an optional argument in parentheses in cases where the 
%% data center identification differs from what is to be printed in the paper.

% -ISM: kinematics and dynamics
% can't say there's a whole lot of dynamics involved here...

%%%%%%%%%%%%%%%%%%%%%
%                   %
%     Section 1     %
%                   %
%%%%%%%%%%%%%%%%%%%%%
\section{Introduction}
The formation of massive star clusters is a topic of active debate and study \citep[][and references therein]{LK14}.  Current questions include the timescale and mechanisms of gas mass assembly and star formation \citep{pf13,b18}, the degree to which gravity, turbulence, or magnetic fields control the dynamics \citep{c12,zv14,pp16,kk18} and the fidelity with which we can measure these effects with only trace constituents \citep{p18}.

BYF\,73 \citep[= G286.21+0.17, part of the Galactic Census of High and Medium-mass Protostars (CHaMP) survey of molecular clouds;][]{b10,b11} is one of only a dozen or so known parsec-scale molecular clumps that are undergoing large-scale collapse/contraction, but where only a few protostars have formed so far, and the cloud is still gas-dominated \citep{pf13,r13,w16}.  With an estimated mass of 2$\times$10$^4$~$M$\solar\ and luminosity of 10$^4$~$L$\solar, BYF\,73 has the highest measured mass inflow rate, 0.034\,$M$\solar~yr$^{-1}$, even among this extreme cohort \citep{b10,b16}.  Therefore, it may be in the early stages of forming a super star cluster like NGC\,3603 in $\lesssim$0.5\,Myr.

Such gas-dominated clouds are highly significant because the physical conditions, dynamics, and evolution must still be close to the cloud's initial state, as opposed to even slightly more evolved objects like hot cores \citep{gbs14}, where the internal conditions are already dramatically altered by the energy input from luminous protostars.  Therefore, a careful study of each such cloud will provide important boundary conditions for star formation theory.  In this Letter, we present a range of new and archival data on BYF\,73 from $\mu$m to mm wavelengths, in order to examine the embedded protostellar content, gas mass distribution, and excitation conditions.%  In \S2 we briefly describe each data set and its reduction, analyze these results in \S3, and conclude with some broader implications in \S4.
% I don't think an outline is needed for a paper this short

%%%%%%%%%%%%%%%%%%%%%
%                   %
%     Section 2     %
%                   %
%%%%%%%%%%%%%%%%%%%%%
\section{Observations and Data Reduction}

\subsection{Spitzer/IRAC}

\begin{figure}[tb]
\vspace{0mm}
\centering
\includegraphics[angle=-90,scale=0.35]{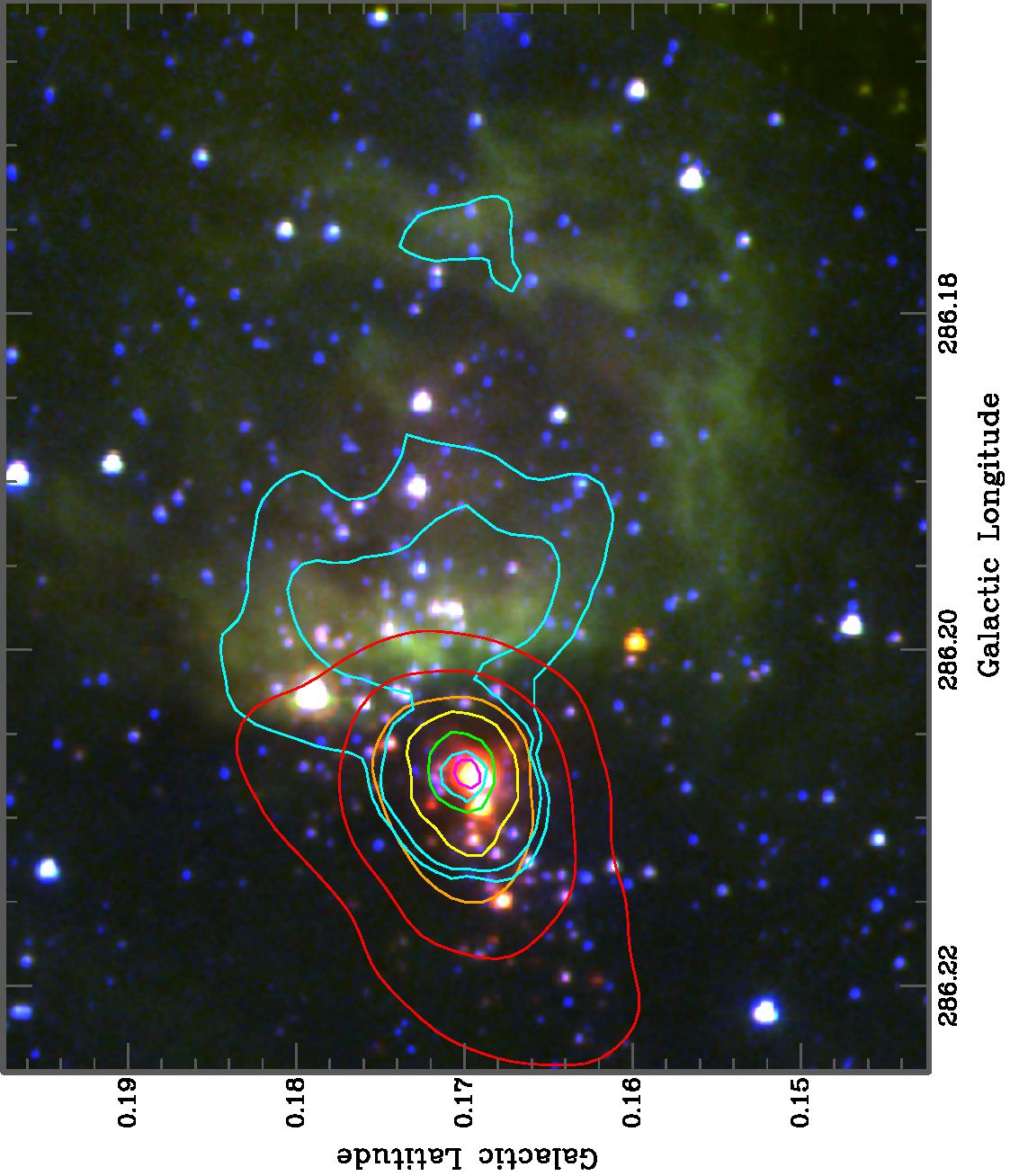}\hspace{1mm}
\vspace{-1mm}
\caption{Composite image of {\em Spitzer} IRAC Bands 2 (red) and 1 (green), with $K$ (blue) from the Anglo-Australian Telescope's IRIS2 camera \citep{b13}.  Contours are overlaid from \emph{Herschel} and Australia Telescope Compact Array (ATCA) data: 500~$\mu$m (red, at 30\% and 50\% of the peak value), 350~$\mu$m (orange, 50\%), 250~$\mu$m (yellow, 50\%), 160~$\mu$m (green, 50\%), 70~$\mu$m (cyan, 2\%, 3\%, 50\%), and 3mm (magenta, 67\%).  Note the clear separation between the molecular cloud to the left, and the compact H\textsc{ii} region to the center and right.
}
\label{irac}\vspace{-2mm}
\end{figure}

BYF\,73 was observed by the {\em Spitzer} Space Telescope in all four IRAC bands (3.6--8~$\mu$m) as part of the Galactic Legacy Infrared Midplane Extraordinaire (GLIMPSE) survey \citep{bc03}.  We downloaded calibrated public data from NASA's IRSA website, and transformed to Galactic coordinates for ease of comparison with data at other wavelengths. We cropped these data to cover all emission associated with the molecular cloud and its adjacent compact H\textsc{ii}~region, as in the $\sim$4$'$ field shown in Fig.\,\ref{irac}.  We measured the background-subtracted flux densities for the eight IRAC stellar sources that were also detected with the Thermal-Region Camera Spectrograph (T-ReCS; \S2.2); these are plotted in Fig.\,\ref{mir-seds}.

%%%%%%%%%%%
%  Fig.1  %
%%%%%%%%%%%
\notetoeditor{}
%\begin{figure*}[ht]
%\end{figure*}

\subsection{Gemini-South/T-ReCS}

\begin{figure}[tb]
\vspace{0mm}
%\centerline{\includegraphics[angle=0,scale=0.38]{byf73-Si2notes.jpg}\hspace{1mm}}
%\centerline{\includegraphics[angle=-90,scale=0.44]{byf73-Q62zoom.jpg}\hspace{1mm}}
\centering
\includegraphics[angle=-90,scale=0.34]{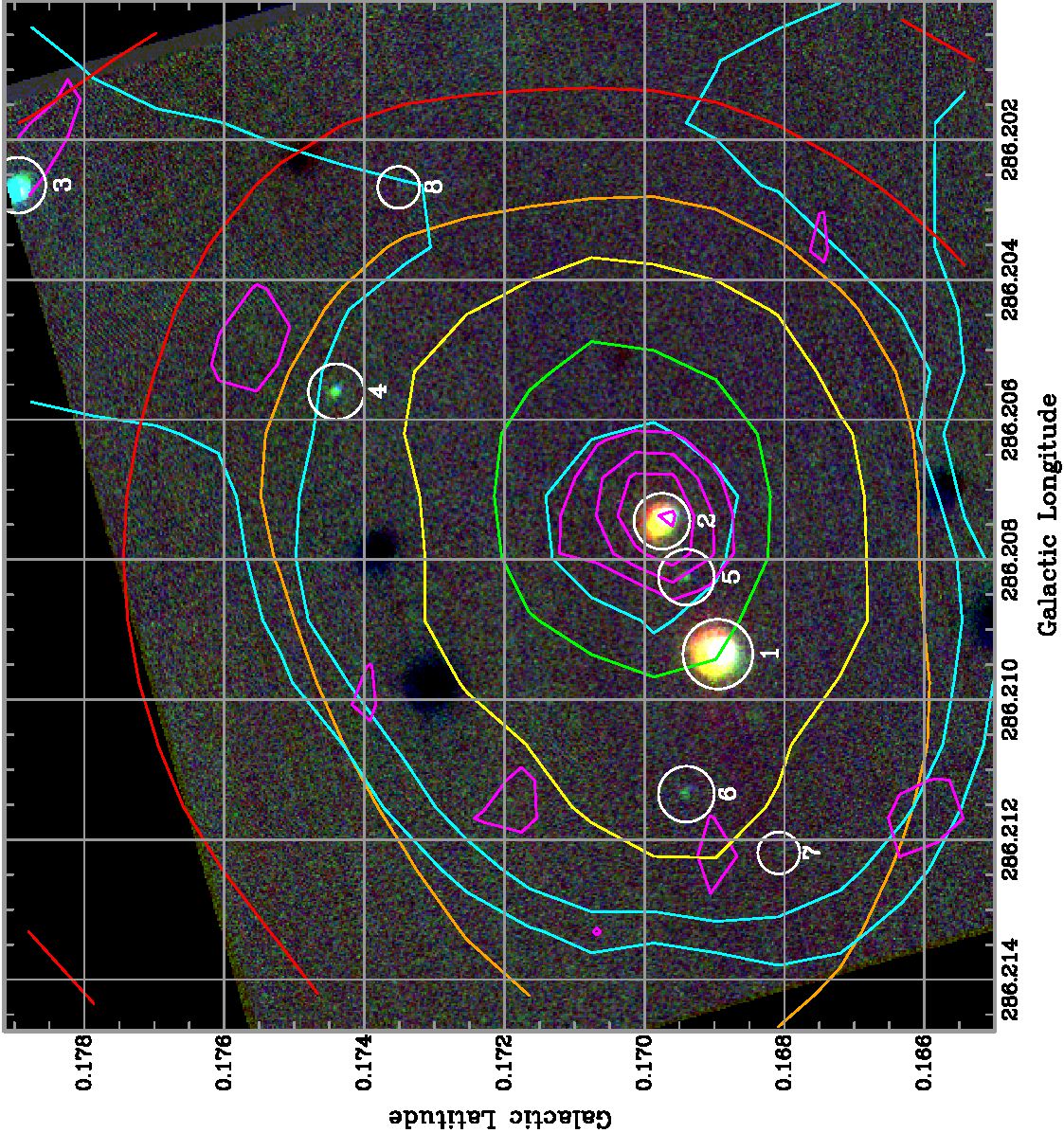}\hspace{1mm}
\vspace{-1mm}
\caption{Identifications of point sources 1--8 detected in part of the $\sim1'$ wide Gemini-South/T-ReCS mosaic.  This composite image has colors (Qa = red, Si6 = green, Si2 = blue) indicating true flux ratio differences between these bands, and is overlaid by ATCA 3mm continuum contours (magenta, from 14\,mJy/bm spaced by 6\,mJy/bm = 1$\sigma$) and the same far-infrared (FIR) contours as in Fig.\,\ref{irac}.  The ``dark'' sources are negative images of the bright ones, due to the 15$''$ chop-throw between the target and reference positions in the telescope observing mode.  Sources MIR\,1, 2, and 5--7 were imaged in all three positions, and so appear as bright objects, flanked 15$''$ to the N and S by dark images.  Sources MIR 4, 8, and 3 (barely) were imaged in the target and one reference position.}
\label{trecs}\vspace{0mm}
\vspace{1mm}
\end{figure}

\begin{figure}[tb]
\centering\includegraphics[width=0.85\columnwidth]{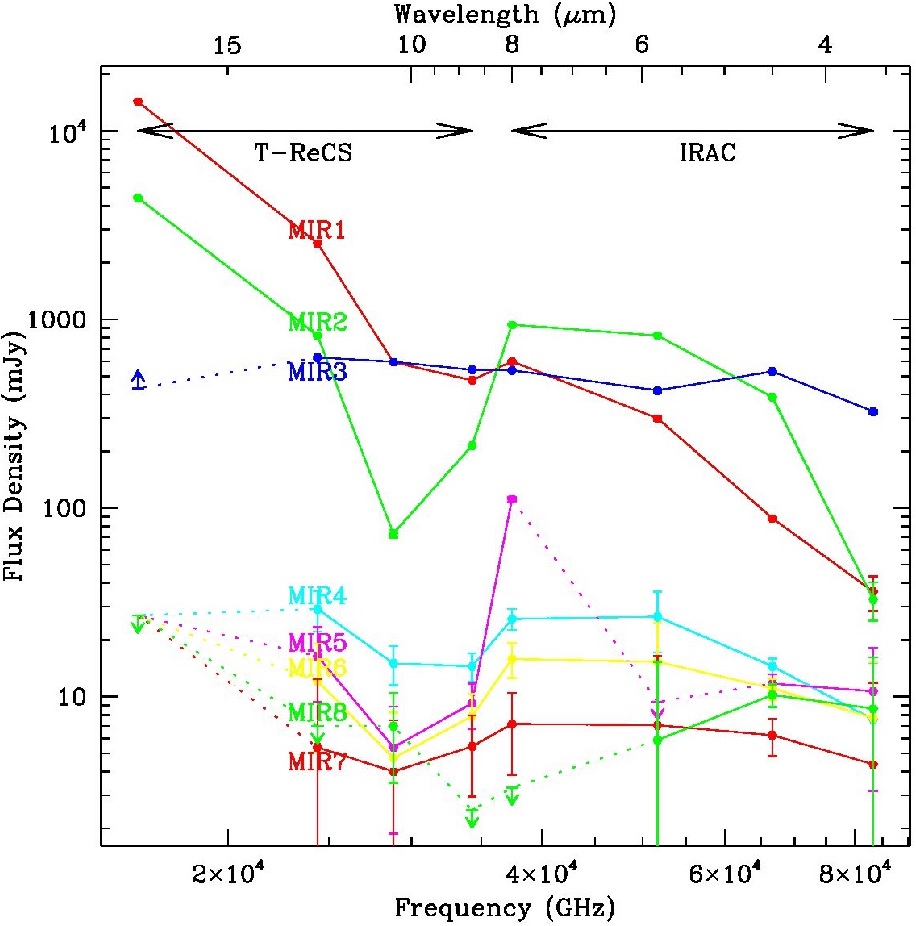}\hspace{1mm}
\vspace{0mm}
\caption{Flux densities of T-ReCS-detected sources in both T-ReCS and {\em Spitzer}/IRAC data.  Secure detections are indicated by 
% SR: dots ->
points connected by solid lines; 3$\sigma$ upper limits are indicated by symbols connected to the other data by dotted lines.  As we obtained only a partial image for MIR 3 at 18.3\,\micron, its flux density is at least twice the indicated lower limit.  Error bars are 1$\sigma$ noise values, but the calibration uncertainty is $\pm$15\% for these data after allowing for subtraction of variable background emission.
}
\label{mir-seds}\vspace{-1mm}
\end{figure}

We obtained deep mid-infrared (MIR) data on BYF\,73 with the high-resolution (point-spread function (PSF) $\approx$ 0\farcs3) T-ReCS camera \citep{t98} at Gemini-South on UT 2010 June $5-6$, as part of program GS-2010A-Q-42 (PI: P. J. Barnes).  The observations were set up to image a 1$'$ field of view with a six-field mosaic of the 20$''$$\times$30$''$ T-ReCS detector area, oriented to capture as many of the sources visible in the {\em Spitzer} images as possible.  We cycled through the four filters Si2 (effective wavelength 8.74\,$\mu$m), Si4 (10.4\,$\mu$m), Si6 (12.3\,$\mu$m), and Qa (18.3\,$\mu$m) during the observing to form a commensurate set of images and enable multi-band photometry of all sources.  \citet{c99} MIR standard stars SAO~250905 and SAO~222647 were observed before and after BYF\,73, respectively, for flux calibration. The data were reduced with an in-house MIR data reduction package, which performed the chop-and-nod correction and co-added frames to form one image of BYF\,73 at each MIR wavelength. Reduced and flux-calibrated images are presented in Fig.\,\ref{trecs} as a composite image of three of the four T-ReCS bands.  The flux densities of the eight detected sources (based on aperture photometry with radii equal to three times the FWHM of the PSF) are included in the spectral energy distribution (SED) plot of Fig.\,\ref{mir-seds}.  For MIR\,1, these are totals for both components of an equal-brightness binary, separated by 0\farcs38 (PA = --14\degree) in the Si2 image, or 950\,au assuming a distance of 2.5\,kpc \citep{b10}. The other MIR sources show no evidence of binarity at this resolution.

%%%%%%%%%%%
%  Fig.2  %
%%%%%%%%%%%
\notetoeditor{}
%\begin{figure*}[ht]
%\end{figure*}

%%%%%%%%%%%
%  Fig.3  %
%%%%%%%%%%%
\notetoeditor{}
%\begin{figure*}[ht]
% {\color{red} SR: How can we be sure these MIR sources are all stars?  PB: Well, they're point sources in a v.small PSF, so they're at least (proto)stars, as much as anything is in the NIR or optical.}
%0''.378056 = 945.1395 au, 4x1 pixels apart
%%% PSFs on MIR 3 %%%
% Si2 = 3 pixels = 0.''2654
% Si4 = 3.5 pix = 0.''3096
% Si6 = 4 x 3.5 p = 0.''47x0.''39

%%%%%%%%%%%
%  Fig.4  %
%%%%%%%%%%%
\begin{figure*}[ht]
\vspace{0mm}
\includegraphics[angle=0,scale=0.132]{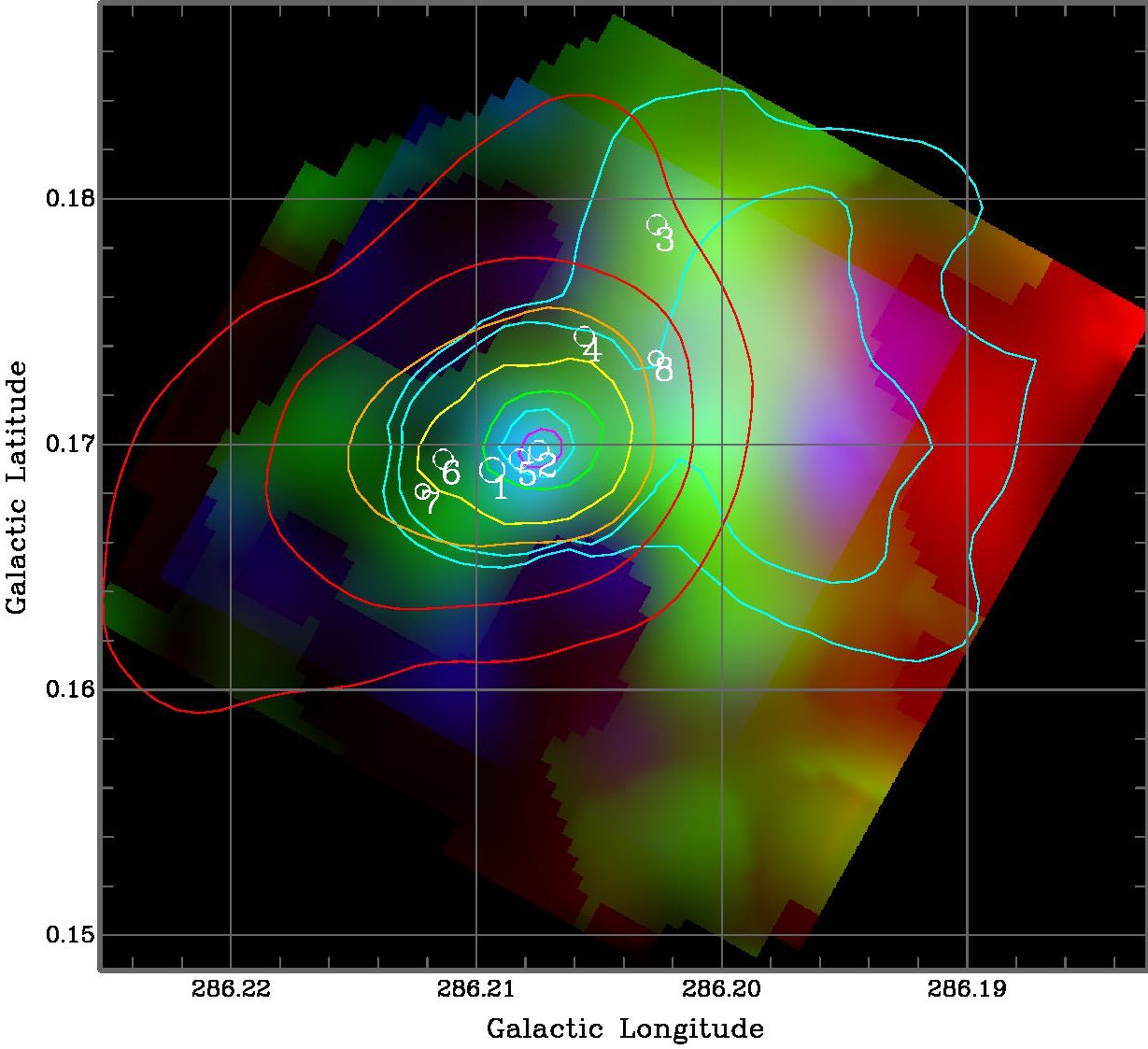}
\includegraphics[angle=0,scale=0.132]{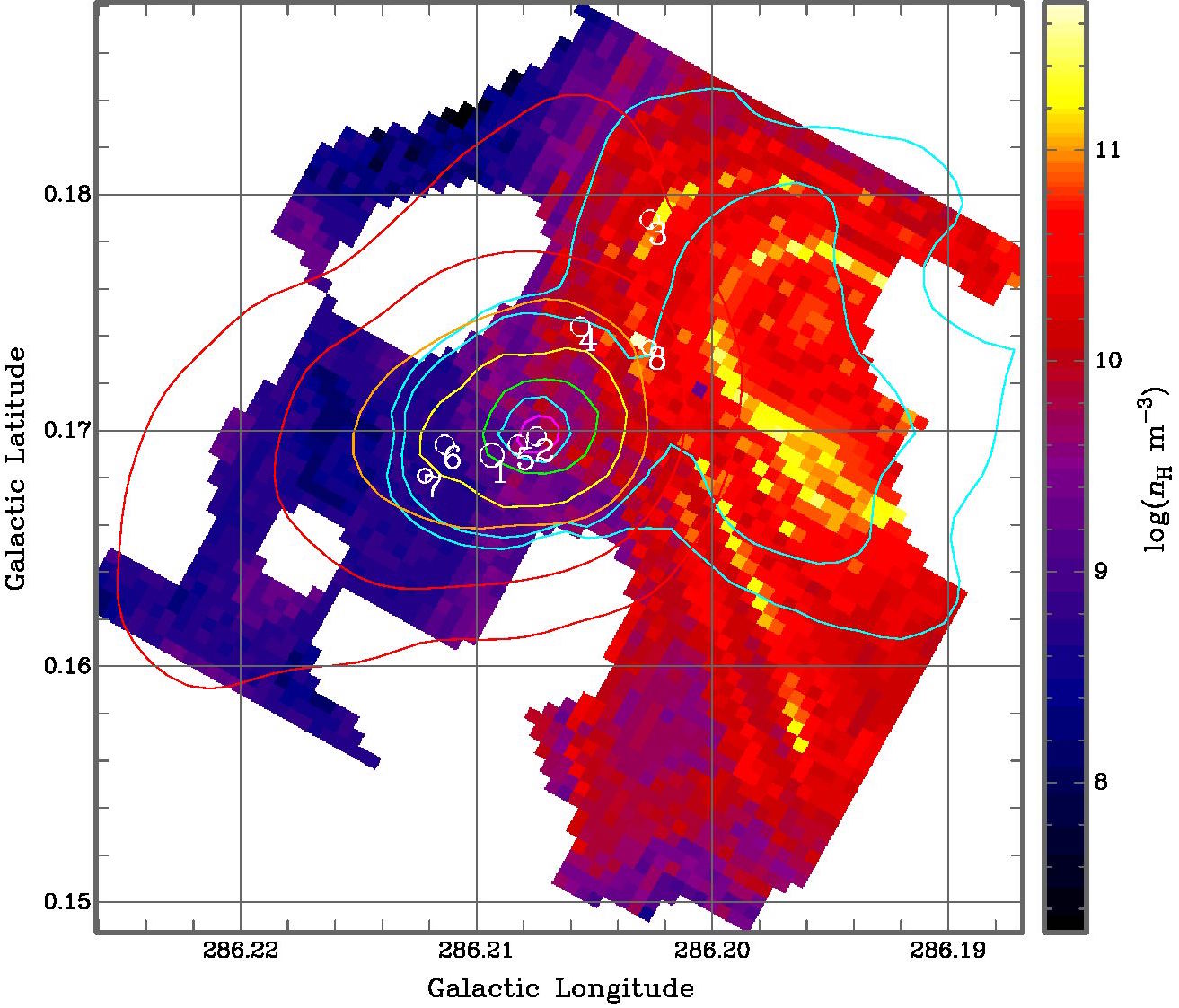}\hspace{0.5mm}

\vspace{-55mm}\hspace{120mm}\includegraphics[angle=-90,scale=0.275]{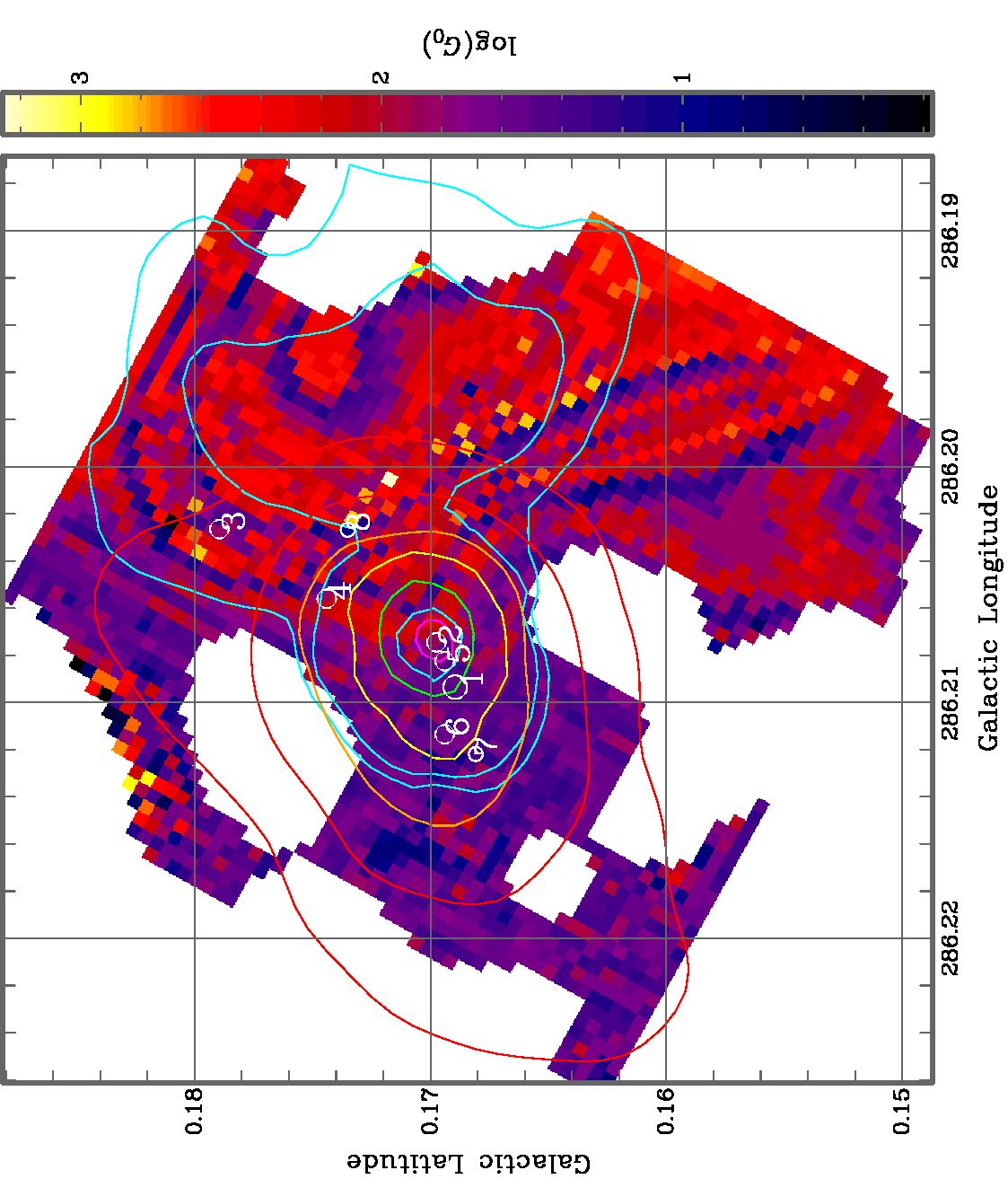}
\vspace{-1mm}
\caption{({\em Left}) Composite image of \cofe\ intensity (red, normalized to 3.8$\times$10$^{-7}$\,\wsqmsr), \ooff\ (green, 6.4$\times$10$^{-8}$\,\wsqmsr), and \ost\ (blue, 9.4$\times$10$^{-11}$\,\wsqmsr).  ({\em Middle} and {\em right}) Respective images of derived log($n_\text{H}$) (m$^{-3}$) and log($G/G_0$) ($G_0$ in \citealt{h68} units, 1.6$\times$10$^{-6}$\,W\,m$^{-2}$), computed by localizing FIFI-LS line intensities/ratios on the PDR Toolbox parameters.  Contours in all panels are the same as in Fig.\,\ref{irac}.
% SR:{\color{red} %What do the contours in each panel represent? The left-hand axes of the middle and right panels should be labeled as "Galactic Latitude", otherwise one is misled into thinking the right panel vertical axis shows log(n$_{H}$)!
% PJB: I was trying to save space... now fixed, and with updated parameter fits from RP.}
}
\label{ratsnwebs}\vspace{0mm}
\end{figure*}

\subsection{Stratospheric Observatory for Infrared Astronomy/Far-Infrared Field-Imaging Line Spectrometer (SOFIA/FIFI-LS)}

We obtained integral-field spectra of BYF\,73 centered on the \ost, \oee, \ooff, and \cofe\ lines on 2016 July 1 \& 3 UT with FIFI-LS \citep{cfg12,kbb14} on board SOFIA, as part of project 04-0061 (PI: P. J. Barnes).
%The [OI]63$\mu$m and [OIII]88$\mu$m lines and continua are observed through the Blue detector channel (FWHM = $6"$, 30$"$ FOV, 50$\mu$m$<\lambda<$125$\mu$m), and the [OI]145$\mu$m and [CII]158$\mu$m lines and continua are observed by the Red channel detector (FWHM = $12"$, 1$'$ FOV, 105$\mu$m$<\lambda<$200$\mu$m)\footnote{\textit{Guest Investigator Handbook for FIFI-LS Data Products}, https://www.sofia.usra.edu/sites/default/files/FIFI-LS\_GI\_Handbook\_RevB1.pdf}.  FIFI-LS's velocity resolutions at the centers of the [OI]63$\mu$m, [OIII]88$\mu$m, [OI]145$\mu$m, and [CII]158$\mu$m lines are $\sim$250~km~s$^{-1}$, $\sim$450~km~s$^{-1}$, $\sim$300~km~s$^{-1}$, and $\sim$270~km~s$^{-1}$, respectively, so no information could be gleaned about the local pressure conditions. 
Chopping and nodding were done asymmetrically due to the many nearby MIR sources to the (Galactic) west and south.  Each integral field spectrum combines over 200 exposures dithered in sub-pixel increments to boost the sampling of the final spatially resampled image cube.\footnote{See the FIFI-LS GI handbook at https://www.sofia.usra.edu/ sites/default/files/FIFI-LS\_GI\_Handbook\_RevB1.pdf, and the Cycle 5 SOFIA observer's handbook at https://www.sofia.usra.edu/ science/proposing-and-observing/sofia-observers-handbook-cycle-5/5-instruments-ii-fifi-ls for details of the spectral \& spatial resolutions and observing modes.}  The total integration times were 1659\;s centered on the \cofe\ line, 3287\;s on the \ooff\ line, 1628\;s on the \oee\ line, and 3318\;s on the \ost\ line.  Pipeline processing with \texttt{FLUXER}\footnote{http://www.ciserlohe.de/fluxer/fluxer.html} includes fitting and separation of line and continuum emission components in each band, and telluric correction.

\subsection{\emph{Herschel} Photodetector Array Camera and Spectrometer (PACS) and Spectral and Photometric Imaging REceiver (SPIRE)}

We obtained archival Level 3.5 data from the \emph{Herschel} \citep{hso} satellite's PACS \citep{pwg10} and SPIRE \citep{gaa10} photometers, %with the FIFI-LS continuum data at 145 and 158 $\mu$m 
in order to combine with the other data described herein and fit SEDs across BYF\,73 (see \S3.2).  The \emph{Herschel} data were originally acquired as part of the Carina Nebula Complex (CNC) open time project \citep[see][for %observational and data reduction 
details]{prg12}. %\texttt{OT1\_tpreibis\_1} which mapped the 

\subsection{Atacama Submillimeter Telescope Experiment (ASTE) and ATCA} \label{sec:atca}

The flux density at 850\,$\mu$m, 8$\pm$1\,Jy, is from a single-pointing measurement at the line emission peak, made with the 10\,m ASTE \citep{ekk04} as part of another project (Y.\,Yonekura 2016, private communication).

We observed BYF\,73 at ATCA on 2010 October 1 and 4 UT in the H75 array (baselines $\approx$ 31--89\,m) in both the 94$\pm$2\,GHz continuum and the \hcop\ and \httco\ \joz\ emission lines (89.19\,GHz and 86.75\,GHz, respectively) as part of program C2288 (PI: Barnes). We used the quasars 0537--441 and 1045--62, respectively, as passband and complex gain calibrators, and Mars as the flux calibrator.  We mapped an 80-pointing mosaic of size 3\farcm2 centered on the peak molecular line emission as measured in the Mopra maps \citep{b10}.  Mediocre weather and poor phase stability, however, challenged the normal \textsc{Miriad} data reduction pipeline \citep{s96}, resulting in line maps with low signal-to-noise ratios (S/N).  The low spectral-line sensitivity was exacerbated by an apparently smooth intrinsic emission structure, as the ATCA \hcop\ line flux was $<$30\% of the Mopra single-dish value \citep{b10}, despite the short baselines.  We fared better in the continuum, clearly detecting the point source MIR\,2 at flux density 34$\pm$7~mJy in the 5\farcs6$\times$4\farcs7 synthesized beam (Fig.\,\ref{trecs}).  MIR\,3 may also have been detected, but at $\sim$2$\sigma$ this detection is not reliable; MIR\,1 was not detected at all.  Deconvolving MIR\,2's measured size, we obtain 4\farcs2$\times$3\farcs$0 = 10,600\times7400$\,au at 2.5\,kpc for its physical dimensions at 3\,mm, consistent with MIR\,2 being a massive protostellar core. 
%% 4.228''x2.967'' =  10,571 x 7417\,au at coordinates $\alpha_{2000}$ = 10$^h$ 38$^m$ 32$^s$\hspace{-1mm}.15, $\delta_{2000}$ = --58$^{\circ}$ 19$'$7\farcs9.  This appears to be about 1$''$ N of MIR-2 in the T-ReCS images.
%%% !!! check all the calibrations...

%%%%%%%%%%%%%%%%%%%%%
%                   %
%     Section 3     %
%                   %
%%%%%%%%%%%%%%%%%%%%%
\section{Analysis and Discussion}\label{sec:anlyz}

\subsection{FIR Spectral Lines and Gas Conditions}\label{ssec:fifi}

%%% PJB: some edits inserted below: blue means probably ok text, red means things we should think about and fix.

To analyze the FIFI-LS data, we localized the overlap of contours of two or more line ratios/intensities on log($n_\text{H}$)--log($G/G_0$) parameter maps from the Photo Dissociation Region Toolbox\footnote{http://dustem.astro.umd.edu/pdrt/models1.html} \citep[PDRT;][]{kwh06,pw08}.  Fig.\,\ref{ratsnwebs} shows the observed line intensity and derived log($n_\text{H}$) and log($G/G_0$) maps, based on the PDRT and observed \cofe\ and \ooff\ line ratios.  No line components were detected in \oee. The \ost\ fluxes were difficult to reconcile with the fluxes in the \ooff\ and \cofe\ lines, likely exacerbated by the \ost\ line's poor separation from the continuum in both the pipeline-processed spectra and our own alternative attempts. %; the core of the line {\color{blue}was badly affected by} telluric correction in later pipeline versions.  
The \ooff/\cofe\ ratio and integrated \cofe\ line flux were enough to determine $n$ and $G/G_0$ along the PDR to about half a dex precision, as shown in Fig.\,\ref{ratsnwebs}.  There, we derive log($G/G_0$) $\approx$ 1.5--2 and $n_\text{H}$ ranging over 10$^{10-11}$\,m$^{-3}$.

Outside of the PDR, especially near MIR\,1--2, PDRT's built-in assumption of $A_V\leq10^m$ breaks down, and there are no prescriptions for higher extinctions. $G/G_0$ did not noticeably change near MIR\,1-3, but $n$ fell where it was expected to rise. We found that the combination of weak \cofe\ line emission (at levels PDRT that flags as unreliable) and moderately high \ooff/\cofe\ ratios around MIR\,1-3 create contours in the PDRT maps of $n$ and $G/G_0$ that are nearly parallel over about three orders of magnitude in $n$ (Fig.\,\ref{ratsnwebs}).

While the \ost/\ooff\ ratios do not provide further useful constraints on $n$ or $G/G_0$ at most locations, near MIR\,2 there is a suggestion of a distinct high-density peak in the gas (10$^{10.7}$\;m$^{-3}$, not shown here), albeit with large uncertainty ($\sim$3 orders of magnitude in log\,$n$).  This is reflected in the slightly better-constrained three-line localisation shown in Fig.\,\ref{ratsnwebs}, where log\,$n = 9.5^{4.25}_{-1}$ at MIR\,2. %in Fig.\,\ref{fullsed}a.  
To check this, we determined the mass (and thus the density) and luminosity of MIR\,2 from SED fitting (\S\ref{ssec:seds}).  Assuming MIR\,2 is an ellipsoid with a line of sight depth similar to its observable dimensions ($\sim$9000\,au), its volumetric mean density is $n$ = $\rho/(\mu m_{\rm H})\approx 8\times$10$^{13}$~m$^{-3}$ (\S\ref{ssec:seds}), close to PDRT's mean+1$\sigma$ value of $n$ given the temperature and $G/G_0$, and lending credence to the PDRT result despite its large uncertainty.

At these densities, in a relatively weak far-ultraviolet (FUV) field, and given that BYF\,73 is on average cool and dense enough to have a CO-depleted center \citep[][subm.]{p18}, carbon should be locked up in CO.  The measured line luminosities around MIR\,2 are $\sim$7$\times$10$^{-3}$\,$L$\solar in the \cofe\ line and $\sim$4$\times$10$^{-3}$\,$L$\solar in the \ooff\ line.  We conclude that in BYF\,73, FIR ``cooling'' lines of O and C do not contribute significantly to the energy balance of the cloud, compared to the total FIR luminosity (\S3.3).

\subsection{MIR Point Sources}\label{ssec:mir}

The MIR photometry reveals three features. First, we detect only six (proto)stars near the center of the molecular cloud, despite deep, high-resolution Very Large Telescope (VLT) $JHK$ data showing $\sim$30 heavily reddened objects ($J$--$K$ $>$ 3) in the same 30$''$-wide area \citep{a17}.  The other two MIR stars, MIR\,3 and 8, are either located within or projected onto the PDR front, and so may be slightly more evolved young stellar objects.  The six central MIR stars may be the only true protostars in the imaged area, while the remaining near-infrared (NIR) objects may just be foreground pre-main sequence stars.

%%%%%%%%%%
% RLP: This paragraph needs to be reevaluated, as the ref objected to all but the first sentence. I don't think we can really distinguish Class 0 and I with the data we have.  I also need to look for some modernized version of  http://adsabs.harvard.edu/abs/1994ASPC...65..197B

%% PJB: I tried to put in some appropriate edits.
Second, only MIR\,1--3 contribute significantly to the bolometric luminosity of the cloud.  MIR\,3--8 have slowly rising SEDs in the MIR, resembling classic Class II or III protostars (\citealt{als87,b94,am94}; MIR\,5 seems to be brighter than expected in the IRAC band 4 image, but is the most affected by blending with MIR\,2, so this data point should be treated with caution.)  Only the SEDs of MIR\,1 and 2 show the steep rise at longer wavelengths expected of Class 0 or I protostars, although MIR\,1 will need higher-resolution FIR data to separate it cleanly from its much brighter neighbor MIR\,2.  MIR\,2 seems to fall between Class 0 and I definitions: it has a relatively high $T_{\rm bol}$ and low $L_{\rm submm}/L_{\rm bol}$, suggesting Class I, but very high gas fraction and infall rate, plus small age $t_{\rm inf}=M/\dot{M}\sim7000$\,yr, suggesting Class 0.
%%%%%%%%%%%%

Third, all of the MIR stars except MIR\,3 and 8 (the two stars possibly within the PDR) show absorption near 10\,$\mu$m, attributable to the 9.7\,$\mu$m silicate feature and indicating the presence of intervening or circumstellar cold dust.  For MIR\,1 and 4--7, the absorption has a depth to 30--50\% of the adjacent 12\,$\mu$m or 8\,$\mu$m continuum, indicating dust with optical depths at 10\,$\mu$m near 1.  For MIR\,2, however, the absorption is $>$90\% of the MIR continuum, suggesting correspondingly larger amounts of dust, $\tau_{10}\sim 3 \pm1$.  % other numbers also tweaked
Using a conversion of $A_{V}$/$\tau_{9.7}$ = 18.5$\pm$1.0 magnitudes \citep{m90}, we find approximate values of visual extinction toward each star of 54$\pm$18$^m$ (MIR\,2) or 18$\pm$5$^m$ (others), though these conversions are for diffuse dust.  In molecular clouds, this conversion shows substantial variation---often flattening at large $A_V$ \citep{cep07}---so these extinction estimates are likely lower limits.  Assuming further standard conversions of $N_{\rm H}$ = 1.87$\times$10$^{25}$\,m$^{-2}$\,$A_V$ $\approx$ 2\nhtwo\ in these clouds \citep{gl17}, where the H\textsc{i} contribution to the total column density is assumed to be small, $\Sigma = 1.88$\,$M$\solar\,pc$^{-2}(N_{\text{H}_2}/10^{24} m^{-2})$\citep{b18}, and multiplying by 2 to account for the rear half of each core, we obtain corresponding (very approximate) total mass columns from the silicate absorption of $1950 \pm 650$\,$M$\solar\,pc$^{-2}$ (MIR\,2) and $650 \pm 190$\,$M$\solar\,pc$^{-2}$ (others).  If we assume a fiducial envelope size of 10$^4$\,au \citep{als87}, the total masses (better estimated in \S\ref{ssec:seds}) are $\sim110 \pm 40$\,$M$\solar\,for MIR\,2 and $\sim 40 \pm 10$\,$M$\solar\,for the others.  As we show below, this approach is a case in point for the caution urged by \citet{gl17}.
%3x18.5x18.7/2x1.88x2 = 1951.158
% 18.5x18.7/2x1.88x2 = 650.386
%{\color{red}PJB: We still need to get mass estimates for some other MIR stars. RLP: I put some in further down in the Herschel section. It's kinda hand-wavy, but I hope it'll do.}

\subsection{FIR/Sub-mm Continuum SED Fitting}\label{ssec:seds}

%%%%%%%%%%%
%  Fig.5  %
%%%%%%%%%%%
%\begin{figure*}
\begin{figure}
\includegraphics[width=\columnwidth]{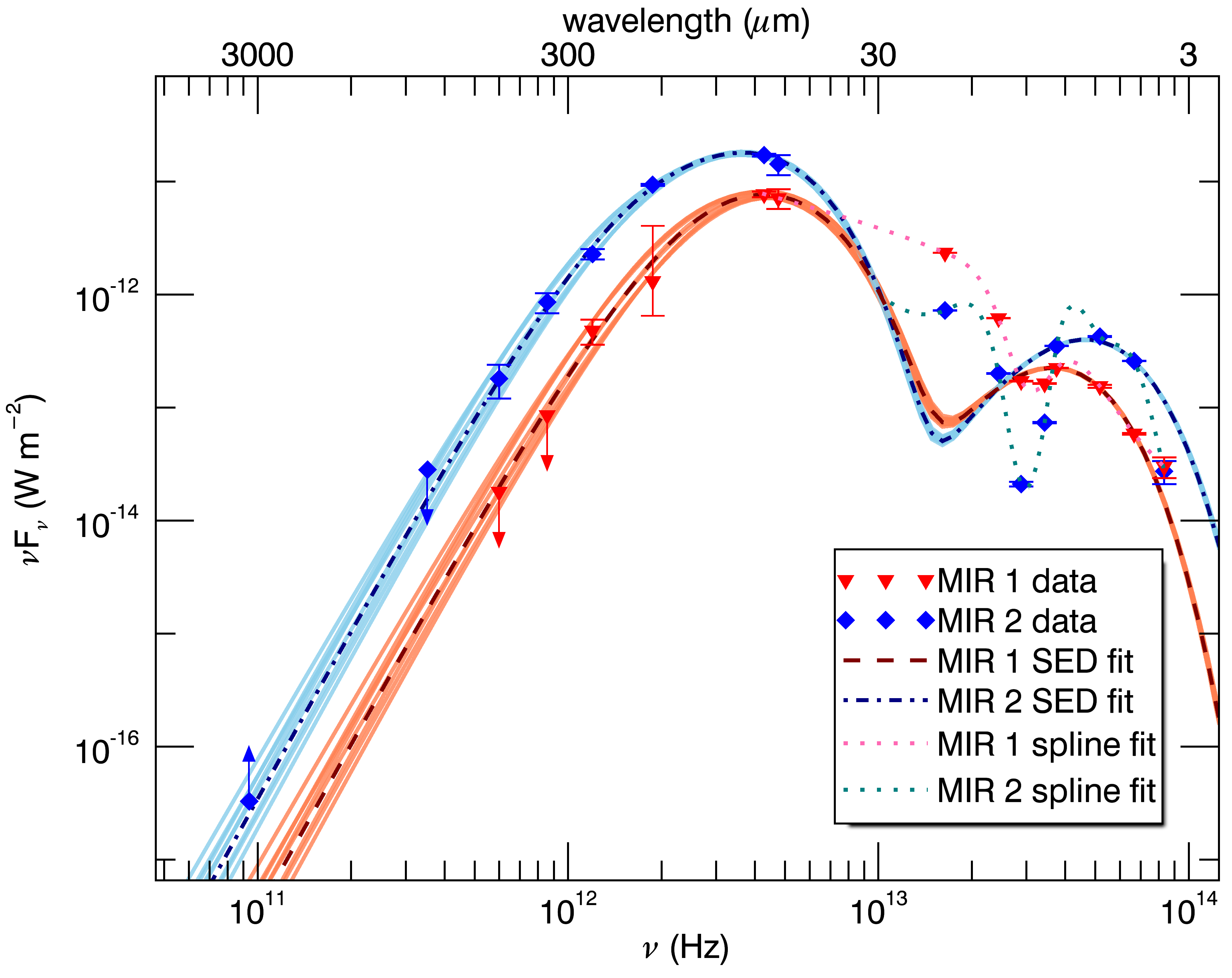}\caption{Two-component SED fits to MIR\,1 and 2. The model used does not fit the 10\,$\mu$m silicate feature or small grain emission.}
\label{fig:fullseds}
\end{figure}

%PJB: Hereafter, I've made numerous small edits to trim the verbiage.
To compute the core masses for the MIR point sources, we cropped and regridded the \emph{Herschel} and SOFIA \ost\ continuum images to the area and pixel grid of the FIFI-LS [C\,{\sc ii}] image, and experimented with fitting up to three 2D Gaussian PSFs at each wavelength, fixed at their respective MIR positions. %of MIR\,1 and MIR\,2.  
MIR\,1 and 2 are 7\farcs4 apart, fortuitously aligned with the minor axes of the PACS beams, so they are separable at 63\,$\mu$m and 70\,$\mu$m, but not at longer wavelengths.
%PJB: I think this footnote is redundant. \footnote{\textcolor{red}{MIR\,1 and 2 become unresolved at $\lambda>97\,\micron$ for \textit{Herschel} and $\lambda>70\,\micron$ for SOFIA.}}  
MIR\,5 is too close to both MIR\,1 and 2 to be separable by \textit{Herschel} or SOFIA, but shorter-wavelength data %(e.g. T-ReCS colors) %%PJB: fluxes?
(Fig.~\ref{mir-seds}) already suggest that MIR\,3--8 contribute negligibly in the FIR, compared to MIR\,1--2; even MIR\,3 is undetected as a resolved object for $\lambda$$>$30\,\micron.\footnote{There is extended FIR/sub-mm structure in the direction of MIR\,6--7, but we doubt either of these sources contributes meaningfully to this emission.}  %%PJB: These sentences are also probably redundant, or too wordy. E.g., MIR\,3, the most isolated and third-brightest MIR object, can in principle be resolved from all the others out to 160~\micron, yet it was undetected at any wavelength longer than 12.3~\micron.  Estimates of the distribution of flux densities at $\lambda\geq160$~\micron\ reflect the assumption that MIR\,1 and 2 are the main point sources.  
At 63\,$\mu$m and 70\,$\mu$m, we fit Gaussians and background levels to minimize the residuals at MIR\,1 and 2.  At longer wavelengths, we %calculated the integrated flux of the multi-Gaussian fit and 
estimated the contribution of MIR\,1 by comparing the shape of the PSF at MIR\,2 on the side facing MIR\,1, to that on the opposite side.  We then integrated the Gaussian models for each core separately at each wavelength, without the background.  %%Fits to MIR\,5 yielded peak fluxes below the background and well below the uncertainties at every wavelength, so we removed it from consideration.  Model fluxes for MIR\,6 were well within the margins of error for MIR\,1 and 2, so it too was excluded.

We fit the resulting fluxes with a modified Planck function of the form $I_{\nu} = F_{\nu}/\Omega \approx B_{\nu}(T_{\text{d}})[1-e^{-\tau}]$ with $\tau = (\nu/\nu_0)^{\beta}N_{\text{H}_2}\mu\,m_{\text{H}}\kappa_0/\gamma$.  Here, $B_{\nu}(T_{\text{d}})$ is the Planck function at dust temperature $T_{\text{d}}$; $\Omega$ is the solid angle of the source; $\beta$ is the dust emissivity index; $\kappa_0$ is the dust opacity coefficient at fiducial frequency $\nu_0$ (we use 1200 GHz); $N_{\text{H}_2}$ is the fitted H$_2$ column density; $\gamma$ is the gas-to-dust mass ratio; $\mu$=2.8 is the mean molecular weight per hydrogen molecule; and $m_\text{H}$ is the mass of the hydrogen atom.  $T_{\text{d}}$ and $N_{\text{H}_2}$ were allowed to vary, while $\gamma$ was fixed at 100 \citep{bsc90}, $\beta$ at 2.0 \citep[][and references therein]{bianchi99}, and $\kappa_0=\kappa(350\micron)\approx$ 0.2\,m$^2$\,kg$^{-1}$ \citep[e.g.,][]{LLL15}.  For MIR\,2, we used $\Omega$ from the ATCA interferometry, 14.3~arcsec$^2$ (see \S\ref{sec:atca} for dimensions), as a representative core size ($D^2\Omega=2.09\times10^{-3}$\,pc$^2$ where $D=2.5$\,kpc).  MIR\,1 is only resolved in the IRAC 3.6\,$\mu$m band, so we estimated MIR\,1's deconvolved core dimensions to be 1\farcs$9\times1$\farcs6 ($D^2\Omega=5.1\times10^{-4}$\,pc$^2$), noting that the IRAC 3.6\,$\mu$m filter passband encloses the 3.29\,$\mu$m polycyclic aromatic hydrocarbon (PAH) line, which is correlated with dust emission \citep{jones15}.
%1.925"x1.646"

Two-component SED fitting of ATCA, ASTE, {\em Herschel}, SOFIA-63$\mu$m, and {\em Spitzer}/IRAC data yields $N_{{\rm H}_2}$ = 6.2$^{+3.6}_{-1.7}\times$10$^{28}$\,m$^{-2}$ in $T_{{\rm d}}$ = 44.8$\pm$0.4\,K gas for MIR\,2, and $N_{{\rm H}_2}$ = 2.1$^{+1.3}_{-0.5}\times$10$^{28}$\,m$^{-2}$ in $T_{{\rm d}}$ = 51$\pm$1\,K gas for MIR\,1 (see Fig.~\ref{fig:fullseds}).  We only included additional warm temperature components to verify that they contributed negligibly to the total mass column and to estimate the total luminosity; their parameters are unreliable.  The formal errors are based on the standard errors of the model fluxes used in fitting and the reported calibration errors of T-ReCS and the {\em Herschel} instruments.  The total uncertainty of $N_{\text{H}_2}$  may be a factor of two or more due to the uncertainty in $\gamma$ (\citealt{rhb15}).  The 9.7~\micron\ silicate absorption feature is not part of the model, and neither is stochastic heating.

For MIR\,2, the above column density corresponds to $\tau_{{\rm MIR}2}(70\,\mu{\rm m})\approx14$, $A_V\sim7000^m$, and $\Sigma_{\text{MIR}2}$ = 1.2$^{+0.3}_{-0.4}\times10^5$ $M$\solar\,pc$^{-2}$, where core the warm component's contribution is negligible.  These values look extreme, but over the representative size of MIR\,2 at 3\,mm, the core mass works out to a reasonable 240$^{+80}_{-50}$\,$M$\solar.  The total luminosity of MIR\,2 is 4700$^{+100}_{-500}$\;$L$\solar, of which about 4500\;$L$\solar\ is from the $T_{{\rm d}}=45$\,K component alone. Despite being only a few percent of BYF\,73's total mass and not much warmer than the clump on average \citep{p18}, MIR\,2 contributes nearly half the total luminosity.  Deepening the mystery, if $\dot{M}=0.034\pm0.017$\,$M$\solar\,yr$^{-1}$ \citep{b10} and $R \approx 4500$\,au, the gravitational contraction luminosity is $L_{\rm g}=GM\dot{M}R^{-1}\approx260$\,$L$\solar, much more than the line luminosity but still only $\sim$6\% of MIR\,2's total.

For MIR\,1, by the same analysis, $\tau_{{\rm MIR}1}(70\,\mu{\rm m})\sim5$ and $\Sigma_{{\rm MIR}1}$ $\approx$ 4$^{+3}_{-2}\times10^4$\,$M$\solar\,pc$^{-2}$, corresponding to a core mass of about 20$^{+20}_{-10}$\,$M$\solar.  The cool dust component of MIR\,1 contributes $1100 \pm 100$\,$L$\solar\ to the total luminosity. The T-ReCS 12\,$\mu$m and 18\,$\mu$m data could only be included in the fit with a spline (which was only used to estimate the total luminosity), but this raised MIR\,1's total luminosity to $\sim1900$\,$L$\solar.  These T-ReCS data points cannot trace the same temperature component as the FIR data, because that would indicate a 1\,$M$\solar\ protostar with a 4000\;$L$\solar\ luminosity.

For MIR\,3, given the background flux levels in the \emph{Herschel} data and its non-detection at $\lambda\geq30$\,\micron, MIR\,3 can be no more than $\sim$10\% of the mass of MIR\,1. If the same holds for MIR\,4--8 as indicated by the FIR non-detections, MIR\,2 has nearly 10 times the mass of the other seven objects put together.

%%%%%%%%%%%%%%%%%%%%%
%                   %
%     Section 4     %
%                   %
%%%%%%%%%%%%%%%%%%%%%
\section{Concluding Remarks}
The mass and volume density of MIR\,2 is comparable to the larger of two massive cores in SDC\,335 \citep{pf13}, a similar molecular clump to BYF\,73 undergoing a massive inflow of gas toward its central objects, but in the case of SDC\,335 through a prominent network of accreting filaments.  BYF\,73 is different because the inflow observed in the single-dish data (resolution 40$''$) is across a cloud structure smooth enough to start resolving out on scales $\lesssim$30$''$ $\approx\frac{1}{3}$\,pc.  Furthermore, while the two cores in SDC\,335 account for about 10\% of the total mass of that cloud, MIR\,2 is potentially $\sim10\times$ as massive as MIR\,1 and MIR\,3--8 combined, yet accounts for only about 1\% of BYF\,73's total mass: it is $\gtrsim$98\% gas.  This result means BYF\,73 may represent an even earlier stage of massive star formation than SDC\,335, such that much of the cloud still shows signs of CO freeze-out and has detectable sublimation fronts facing both NGC\,3324 and the adjacent compact H\,{\sc ii} region \citep{p18}.  In the hunt for the elusive transition from starless core to massive Class 0 protostar, the cores in BYF\,73 may be the closest yet seen. % BYF\,73's larger size could also mean that its evolutionary stage is similar to SDS335's, but its gravity has helped it retain and insulate more of its original mass

%%%%%%%%%%%%%%%
%             %
%  Section 5  %
%             %
%%%%%%%%%%%%%%%

%% No more than seven \figcaption commands are allowed per page,
%% so if you have more than seven captions, insert a \clearpage
%% after every seventh one.

\acknowledgments

We thank the SOFIA crew and Gemini-South \& {\em Spitzer} staff for outstanding support of their respective telescopes, Vicki Lowe for help with the ATCA observing, and the anonymous referee for several helpful suggestions which improved the paper.  R.L.P. and P.J.B. gratefully acknowledge support from grants NASA-ADAP NNX15AF64G and SOF 04-0061.  %SDR ... DL ... 
% SR: Standard ATNF acknowledgement from
% https://www.atnf.csiro.au/research/publications/Acknowledgements.html
The Australia Telescope Compact Array is part of the Australia Telescope National Facility which is funded by the Australian Government for operation as a National Facility managed by CSIRO.
% SR: Standard Gemini acknowledgement from
% https://www.gemini.edu/sciops/data-and-results/acknowledging-gemini
% PB: Ah!  Thanks, Stuart!
Based on observations obtained at the Gemini Observatory, which is operated by the Association of Universities for Research in Astronomy, Inc., under a cooperative agreement with the NSF on behalf of the Gemini partnership: the National Science Foundation (United States), the National Research Council (Canada), CONICYT (Chile), Ministerio de Ciencia, Tecnolog\'{i}a e Innovaci\'{o}n Productiva (Argentina), and Minist\'{e}rio da Ci\^{e}ncia, Tecnologia e Inova\c{c}\~{a}o (Brazil). 

Facilities: \facility{ATCA, SOFIA(FIFI-LS), Gemini:South(T-ReCS), {\em Spitzer}(IRAC), {\em Herschel}(PACS, SPIRE).}

\bibliographystyle{yahapj}
\bibliography{sofia73bib}

\begin{thebibliography}{}
\providecommand\natexlab[1]{#1}
\providecommand\JournalTitle[1]{#1}

\bibitem[{{Adams} {et~al.}(1987){Adams}, {Lada}, \& {Shu}}]{als87}
{Adams}, F.~C., {Lada}, C.~J., \& {Shu}, F.~H. 1987,
  \href{http://dx.doi.org/10.1086/164924}{\JournalTitle{\apj}, 312, 788}

\bibitem[{{Andersen} {et~al.}(2017){Andersen}, {Barnes}, {Tan}, {Kainulainen},
  \& {de Marchi}}]{a17}
{Andersen}, M., {Barnes}, P.~J., {Tan}, J.~C., {Kainulainen}, J., \& {de
  Marchi}, G. 2017,
  \href{http://dx.doi.org/10.3847/1538-4357/aa9072}{\JournalTitle{\apj}, 850,
  12}

\bibitem[{{Andre} \& {Montmerle}(1994)}]{am94}
{Andre}, P., \& {Montmerle}, T. 1994,
  \href{http://dx.doi.org/10.1086/173608}{\JournalTitle{\apj}, 420, 837}

\bibitem[{{Barnes} {et~al.}(2018){Barnes}, {Hernandez}, {Muller}, \&
  {Pitts}}]{b18}
{Barnes}, P.~J., {Hernandez}, A.~K., {Muller}, E., \& {Pitts}, R.~L. 2018,
  \href{https://www.astro.ufl.edu/~pjb/research/champ/papers/champIV.pdf}{\JournalTitle{\apj},
  to appear}

\bibitem[{{Barnes} {et~al.}(2016){Barnes}, {Hernandez}, {O'Dougherty}, {Schap},
  \& {Muller}}]{b16}
{Barnes}, P.~J., {Hernandez}, A.~K., {O'Dougherty}, S.~N., {Schap}, III, W.~J.,
  \& {Muller}, E. 2016,
  \href{http://dx.doi.org/10.3847/0004-637X/831/1/67}{\JournalTitle{\apj}, 831,
  67}

\bibitem[{{Barnes} {et~al.}(2013){Barnes}, {Ryder}, {O'Dougherty}, {Alvarez},
  {Delgado-Navarro}, {Hopkins}, \& {Tan}}]{b13}
{Barnes}, P.~J., {Ryder}, S.~D., {O'Dougherty}, S.~N., {et~al.} 2013,
  \href{http://dx.doi.org/10.1093/mnras/stt607}{\JournalTitle{\mnras}, 432,
  2231}

\bibitem[{{Barnes} {et~al.}(2010){Barnes}, {Yonekura}, {Ryder}, {Hopkins},
  {Miyamoto}, {Furukawa}, \& {Fukui}}]{b10}
{Barnes}, P.~J., {Yonekura}, Y., {Ryder}, S.~D., {et~al.} 2010,
  \href{http://dx.doi.org/10.1111/j.1365-2966.2009.15890.x}{\JournalTitle{\mnras},
  402, 73}

\bibitem[{{Barnes} {et~al.}(2011){Barnes}, {Yonekura}, {Fukui}, {Miller},
  {M{\"u}hlegger}, {Agars}, {Miyamoto}, {Furukawa}, {Papadopoulos}, {Jones},
  {Hernandez}, {O'Dougherty}, \& {Tan}}]{b11}
{Barnes}, P.~J., {Yonekura}, Y., {Fukui}, Y., {et~al.} 2011,
  \href{http://dx.doi.org/10.1088/0067-0049/196/1/12}{\JournalTitle{\apjs},
  196, 12}

\bibitem[{{Barsony}(1994)}]{b94}
{Barsony}, M. 1994, in Astronomical Society of the Pacific Conference Series,
  Vol.~65, Clouds, Cores, and Low Mass Stars, ed. D.~P. {Clemens} \&
  R.~{Barvainis}, 197

\bibitem[{{Beckwith} {et~al.}(1990){Beckwith}, {Sargent}, {Chini}, \&
  {Guesten}}]{bsc90}
{Beckwith}, S.~V.~W., {Sargent}, A.~I., {Chini}, R.~S., \& {Guesten}, R. 1990,
  \href{http://dx.doi.org/10.1086/115385}{\JournalTitle{\aj}, 99, 924}

\bibitem[{{Benjamin} {et~al.}(2003){Benjamin}, {Churchwell}, {Babler}, {Bania},
  {Clemens}, {Cohen}, {Dickey}, {Indebetouw}, {Jackson}, {Kobulnicky},
  {Lazarian}, {Marston}, {Mathis}, {Meade}, {Seager}, {Stolovy}, {Watson},
  {Whitney}, {Wolff}, \& {Wolfire}}]{bc03}
{Benjamin}, R.~A., {Churchwell}, E., {Babler}, B.~L., {et~al.} 2003,
  \href{http://dx.doi.org/10.1086/376696}{\JournalTitle{\pasp}, 115, 953}

\bibitem[{{Bianchi} {et~al.}(1999){Bianchi}, {Davies}, \& {Alton}}]{bianchi99}
{Bianchi}, S., {Davies}, J.~I., \& {Alton}, P.~B. 1999, \JournalTitle{\aap},
  344, L1

\bibitem[{{Chiar} {et~al.}(2007){Chiar}, {Ennico}, {Pendleton}, {Boogert},
  {Greene}, {Knez}, {Lada}, {Roellig}, {Tielens}, {Werner}, \&
  {Whittet}}]{cep07}
{Chiar}, J.~E., {Ennico}, K., {Pendleton}, Y.~J., {et~al.} 2007,
  \href{http://dx.doi.org/10.1086/521789}{\JournalTitle{\apjl}, 666, L73}

\bibitem[{{Cohen} {et~al.}(1999){Cohen}, {Walker}, {Carter}, {Hammersley},
  {Kidger}, \& {Noguchi}}]{c99}
{Cohen}, M., {Walker}, R.~G., {Carter}, B., {et~al.} 1999,
  \href{http://dx.doi.org/10.1086/300813}{\JournalTitle{\aj}, 117, 1864}

\bibitem[{{Colditz} {et~al.}(2012){Colditz}, {Fumi}, {Geis}, {H{\"o}nle},
  {Klein}, {Krabbe}, {Looney}, {Poglitsch}, {Raab}, {Savage}, {Rebell}, \&
  {Fischer}}]{cfg12}
{Colditz}, S., {Fumi}, F., {Geis}, N., {et~al.} 2012,
  \href{http://dx.doi.org/10.1117/12.924510}{in \procspie, Vol. 8446,
  Ground-based and Airborne Instrumentation for Astronomy IV}, 844617

\bibitem[{{Crutcher}(2012)}]{c12}
{Crutcher}, R.~M. 2012,
  \href{http://dx.doi.org/10.1146/annurev-astro-081811-125514}{\JournalTitle{\araa},
  50, 29}

\bibitem[{{Ezawa} {et~al.}(2004){Ezawa}, {Kawabe}, {Kohno}, \&
  {Yamamoto}}]{ekk04}
{Ezawa}, H., {Kawabe}, R., {Kohno}, K., \& {Yamamoto}, S. 2004,
  \href{http://dx.doi.org/10.1117/12.551391}{in \procspie, Vol. 5489,
  Ground-based Telescopes, ed. J.~M. {Oschmann}, Jr.}, 763

\bibitem[{{Gerin} \& {Liszt}(2017)}]{gl17}
{Gerin}, M., \& {Liszt}, H. 2017,
  \href{http://dx.doi.org/10.1051/0004-6361/201730400}{\JournalTitle{\aap},
  600, A48}

\bibitem[{{Gerner} {et~al.}(2014){Gerner}, {Beuther}, {Semenov}, {Linz},
  {Vasyunina}, {Bihr}, {Shirley}, \& {Henning}}]{gbs14}
{Gerner}, T., {Beuther}, H., {Semenov}, D., {et~al.} 2014,
  \href{http://dx.doi.org/10.1051/0004-6361/201322541}{\JournalTitle{\aap},
  563, A97}

\bibitem[{{Griffin} {et~al.}(2010){Griffin}, {Abergel}, {Abreu}, {Ade},
  {Andr{\'e}}, {Augueres}, {Babbedge}, {Bae}, {Baillie}, {Baluteau}, {Barlow},
  {Bendo}, {Benielli}, {Bock}, {Bonhomme}, {Brisbin}, {Brockley-Blatt},
  {Caldwell}, {Cara}, {Castro-Rodriguez}, {Cerulli}, {Chanial}, {Chen},
  {Clark}, {Clements}, {Clerc}, {Coker}, {Communal}, {Conversi}, {Cox},
  {Crumb}, {Cunningham}, {Daly}, {Davis}, {de Antoni}, {Delderfield}, {Devin},
  {di Giorgio}, {Didschuns}, {Dohlen}, {Donati}, {Dowell}, {Dowell}, {Duband},
  {Dumaye}, {Emery}, {Ferlet}, {Ferrand}, {Fontignie}, {Fox}, {Franceschini},
  {Frerking}, {Fulton}, {Garcia}, {Gastaud}, {Gear}, {Glenn}, {Goizel},
  {Griffin}, {Grundy}, {Guest}, {Guillemet}, {Hargrave}, {Harwit}, {Hastings},
  {Hatziminaoglou}, {Herman}, {Hinde}, {Hristov}, {Huang}, {Imhof}, {Isaak},
  {Israelsson}, {Ivison}, {Jennings}, {Kiernan}, {King}, {Lange}, {Latter},
  {Laurent}, {Laurent}, {Leeks}, {Lellouch}, {Levenson}, {Li}, {Li},
  {Lilienthal}, {Lim}, {Liu}, {Lu}, {Madden}, {Mainetti}, {Marliani}, {McKay},
  {Mercier}, {Molinari}, {Morris}, {Moseley}, {Mulder}, {Mur}, {Naylor},
  {Nguyen}, {O'Halloran}, {Oliver}, {Olofsson}, {Olofsson}, {Orfei}, {Page},
  {Pain}, {Panuzzo}, {Papageorgiou}, {Parks}, {Parr-Burman}, {Pearce},
  {Pearson}, {P{\'e}rez-Fournon}, {Pinsard}, {Pisano}, {Podosek}, {Pohlen},
  {Polehampton}, {Pouliquen}, {Rigopoulou}, {Rizzo}, {Roseboom}, {Roussel},
  {Rowan-Robinson}, {Rownd}, {Saraceno}, {Sauvage}, {Savage}, {Savini},
  {Sawyer}, {Scharmberg}, {Schmitt}, {Schneider}, {Schulz}, {Schwartz},
  {Shafer}, {Shupe}, {Sibthorpe}, {Sidher}, {Smith}, {Smith}, {Smith},
  {Spencer}, {Stobie}, {Sudiwala}, {Sukhatme}, {Surace}, {Stevens}, {Swinyard},
  {Trichas}, {Tourette}, {Triou}, {Tseng}, {Tucker}, {Turner}, {Vaccari},
  {Valtchanov}, {Vigroux}, {Virique}, {Voellmer}, {Walker}, {Ward}, {Waskett},
  {Weilert}, {Wesson}, {White}, {Whitehouse}, {Wilson}, {Winter}, {Woodcraft},
  {Wright}, {Xu}, {Zavagno}, {Zemcov}, {Zhang}, \& {Zonca}}]{gaa10}
{Griffin}, M.~J., {Abergel}, A., {Abreu}, A., {et~al.} 2010,
  \href{http://dx.doi.org/10.1051/0004-6361/201014519}{\JournalTitle{\aap},
  518, L3}

\bibitem[{{Habing}(1968)}]{h68}
{Habing}, H.~J. 1968, \JournalTitle{\bain}, 19, 421

\bibitem[{{Jones} {et~al.}(2015){Jones}, {Bendo}, {Baes}, {Boquien}, {Boselli},
  {De Looze}, {Fritz}, {Galliano}, {Hughes}, {Lebouteiller}, {Lu}, {Madden},
  {R{\'e}my-Ruyer}, {Smith}, {Spinoglio}, \& {Zijlstra}}]{jones15}
{Jones}, A.~G., {Bendo}, G.~J., {Baes}, M., {et~al.} 2015,
  \href{http://dx.doi.org/10.1093/mnras/stu2715}{\JournalTitle{\mnras}, 448,
  168}

\bibitem[{{Kaufman} {et~al.}(2006){Kaufman}, {Wolfire}, \&
  {Hollenbach}}]{kwh06}
{Kaufman}, M.~J., {Wolfire}, M.~G., \& {Hollenbach}, D.~J. 2006,
  \href{http://dx.doi.org/10.1086/503596}{\JournalTitle{\apj}, 644, 283}

\bibitem[{{Klein} {et~al.}(2014){Klein}, {Beckmann}, {Bryant}, {Colditz},
  {Fischer}, {Fumi}, {Geis}, {H{\"o}nle}, {Krabbe}, {Looney}, {Poglitsch},
  {Raab}, {Rebell}, \& {Savage}}]{kbb14}
{Klein}, R., {Beckmann}, S., {Bryant}, A., {et~al.} 2014,
  \href{http://dx.doi.org/10.1117/12.2055371}{in \procspie, Vol. 9147,
  Ground-based and Airborne Instrumentation for Astronomy V}, 91472X

\bibitem[{{Kobayashi} {et~al.}(2018){Kobayashi}, {Kobayashi}, {Inutsuka}, \&
  {Fukui}}]{kk18}
{Kobayashi}, M.~I.~N., {Kobayashi}, H., {Inutsuka}, S.-i., \& {Fukui}, Y. 2018,
  \href{http://dx.doi.org/10.1093/pasj/psy018}{\JournalTitle{\pasj}, 70, S59}

\bibitem[{{Liseau} {et~al.}(2015){Liseau}, {Larsson}, {Lunttila}, {Olberg},
  {Rydbeck}, {Bergman}, {Justtanont}, {Olofsson}, \& {de Vries}}]{LLL15}
{Liseau}, R., {Larsson}, B., {Lunttila}, T., {et~al.} 2015,
  \href{http://dx.doi.org/10.1051/0004-6361/201525641}{\JournalTitle{\aap},
  578, A131}

\bibitem[{{Longmore} {et~al.}(2014){Longmore}, {Kruijssen}, {Bastian}, {Bally},
  {Rathborne}, {Testi}, {Stolte}, {Dale}, {Bressert}, \& {Alves}}]{LK14}
{Longmore}, S.~N., {Kruijssen}, J.~M.~D., {Bastian}, N., {et~al.} 2014,
  \href{http://dx.doi.org/10.2458/azu_uapress_9780816531240-ch013}{\JournalTitle{Protostars
  and Planets VI}, 291}

\bibitem[{{Mathis}(1990)}]{m90}
{Mathis}, J.~S. 1990,
  \href{http://dx.doi.org/10.1146/annurev.aa.28.090190.000345}{\JournalTitle{\araa},
  28, 37}

\bibitem[{{Padoan} {et~al.}(2016){Padoan}, {Pan}, {Haugb{\o}lle}, \&
  {Nordlund}}]{pp16}
{Padoan}, P., {Pan}, L., {Haugb{\o}lle}, T., \& {Nordlund}, {\AA}. 2016,
  \href{http://dx.doi.org/10.3847/0004-637X/822/1/11}{\JournalTitle{\apj}, 822,
  11}

\bibitem[{{Peretto} {et~al.}(2013){Peretto}, {Fuller}, {Duarte-Cabral},
  {Avison}, {Hennebelle}, {Pineda}, {Andr{\'e}}, {Bontemps}, {Motte},
  {Schneider}, \& {Molinari}}]{pf13}
{Peretto}, N., {Fuller}, G.~A., {Duarte-Cabral}, A., {et~al.} 2013,
  \href{http://dx.doi.org/10.1051/0004-6361/201321318}{\JournalTitle{\aap},
  555, A112}

\bibitem[{{Pilbratt} {et~al.}(2010){Pilbratt}, {Riedinger}, {Passvogel},
  {Crone}, {Doyle}, {Gageur}, {Heras}, {Jewell}, {Metcalfe}, {Ott}, \&
  {Schmidt}}]{hso}
{Pilbratt}, G.~L., {Riedinger}, J.~R., {Passvogel}, T., {et~al.} 2010,
  \href{http://dx.doi.org/10.1051/0004-6361/201014759}{\JournalTitle{\aap},
  518, L1}

\bibitem[{{Pitts} {et~al.}(2018){Pitts}, {Barnes}, \& {Varosi}}]{p18}
{Pitts}, R.~L., {Barnes}, P.~J., \& {Varosi}, F. 2018, \JournalTitle{\mnras},
  subm.

\bibitem[{{Poglitsch} {et~al.}(2010){Poglitsch}, {Waelkens}, {Geis},
  {Feuchtgruber}, {Vandenbussche}, {Rodriguez}, {Krause}, {Renotte}, {van
  Hoof}, {Saraceno}, {Cepa}, {Kerschbaum}, {Agn{\`e}se}, {Ali}, {Altieri},
  {Andreani}, {Augueres}, {Balog}, {Barl}, {Bauer}, {Belbachir}, {Benedettini},
  {Billot}, {Boulade}, {Bischof}, {Blommaert}, {Callut}, {Cara}, {Cerulli},
  {Cesarsky}, {Contursi}, {Creten}, {De Meester}, {Doublier}, {Doumayrou},
  {Duband}, {Exter}, {Genzel}, {Gillis}, {Gr{\"o}zinger}, {Henning},
  {Herreros}, {Huygen}, {Inguscio}, {Jakob}, {Jamar}, {Jean}, {de Jong},
  {Katterloher}, {Kiss}, {Klaas}, {Lemke}, {Lutz}, {Madden}, {Marquet},
  {Martignac}, {Mazy}, {Merken}, {Montfort}, {Morbidelli}, {M{\"u}ller},
  {Nielbock}, {Okumura}, {Orfei}, {Ottensamer}, {Pezzuto}, {Popesso},
  {Putzeys}, {Regibo}, {Reveret}, {Royer}, {Sauvage}, {Schreiber}, {Stegmaier},
  {Schmitt}, {Schubert}, {Sturm}, {Thiel}, {Tofani}, {Vavrek}, {Wetzstein},
  {Wieprecht}, \& {Wiezorrek}}]{pwg10}
{Poglitsch}, A., {Waelkens}, C., {Geis}, N., {et~al.} 2010,
  \href{http://dx.doi.org/10.1051/0004-6361/201014535}{\JournalTitle{\aap},
  518, L2}

\bibitem[{{Pound} \& {Wolfire}(2008)}]{pw08}
{Pound}, M.~W., \& {Wolfire}, M.~G. 2008, in Astronomical Society of the
  Pacific Conference Series, Vol. 394, Astronomical Data Analysis Software and
  Systems XVII, ed. R.~W. {Argyle}, P.~S. {Bunclark}, \& J.~R. {Lewis}, 654

\bibitem[{{Preibisch} {et~al.}(2012){Preibisch}, {Roccatagliata}, {Gaczkowski},
  \& {Ratzka}}]{prg12}
{Preibisch}, T., {Roccatagliata}, V., {Gaczkowski}, B., \& {Ratzka}, T. 2012,
  \href{http://dx.doi.org/10.1051/0004-6361/201218851}{\JournalTitle{\aap},
  541, A132}

\bibitem[{{Reach} {et~al.}(2015){Reach}, {Heiles}, \& {Bernard}}]{rhb15}
{Reach}, W.~T., {Heiles}, C., \& {Bernard}, J.-P. 2015,
  \href{http://dx.doi.org/10.1088/0004-637X/811/2/118}{\JournalTitle{\apj},
  811, 118}

\bibitem[{{Rygl} {et~al.}(2013){Rygl}, {Wyrowski}, {Schuller}, \&
  {Menten}}]{r13}
{Rygl}, K.~L.~J., {Wyrowski}, F., {Schuller}, F., \& {Menten}, K.~M. 2013,
  \href{http://dx.doi.org/10.1051/0004-6361/201219574}{\JournalTitle{\aap},
  549, A5}

\bibitem[{{Sault} {et~al.}(1995){Sault}, {Teuben}, \& {Wright}}]{s96}
{Sault}, R.~J., {Teuben}, P.~J., \& {Wright}, M.~C.~H. 1995, in Astronomical
  Society of the Pacific Conference Series, Vol.~77, Astronomical Data Analysis
  Software and Systems IV, ed. R.~A. {Shaw}, H.~E. {Payne}, \& J.~J.~E.
  {Hayes}, 433

\bibitem[{{Telesco} {et~al.}(1998){Telesco}, {Pina}, {Hanna}, {Julian}, {Hon},
  \& {Kisko}}]{t98}
{Telesco}, C.~M., {Pina}, R.~K., {Hanna}, K.~T., {et~al.} 1998,
  \href{http://dx.doi.org/10.1117/12.317279}{in \procspie, Vol. 3354, Infrared
  Astronomical Instrumentation, ed. A.~M. {Fowler}}, 534

\bibitem[{{Wyrowski} {et~al.}(2016){Wyrowski}, {G{\"u}sten}, {Menten},
  {Wiesemeyer}, {Csengeri}, {Heyminck}, {Klein}, {K{\"o}nig}, \&
  {Urquhart}}]{w16}
{Wyrowski}, F., {G{\"u}sten}, R., {Menten}, K.~M., {et~al.} 2016,
  \href{http://dx.doi.org/10.1051/0004-6361/201526361}{\JournalTitle{\aap},
  585, A149}

\bibitem[{{Zamora-Avil{\'e}s} \& {V{\'a}zquez-Semadeni}(2014)}]{zv14}
{Zamora-Avil{\'e}s}, M., \& {V{\'a}zquez-Semadeni}, E. 2014,
  \href{http://dx.doi.org/10.1088/0004-637X/793/2/84}{\JournalTitle{\apj}, 793,
  84}

\end{thebibliography}

\end{document}